# EPR study of some rare-earth ions ($Dy^{3+}$, $Tb^{3+}$ and $Nd^{3+}$) in $YBa_2Cu_3O_6$-compound

M.R. Gafurov[a*], V.A. Ivanshin[a, b], I.N. Kurkin[a], M.P. Rodionova[a], H. Keller[c], M. Gutmann[d], U. Staub[d]

[a]. *MRS Laboratory, Kazan State University, 420008, Kremlevskaya 18, Kazan, Russia*
[b]. *Max-Planck-Institut für Chemische Physik fester Stoffe, 01187 Dresden, Germany*
[c]. *Physik-Institut der Universität Zürich, CH-8057, Zürich, Switzerland*
[d]. *Paul-Sherrer Institut Villigen, CH-5232 Villigen, Switzerland*

E-mail: Marat.Gafurov@ksu.ru

**Abstract**

We investigate the low temperature X-band electron paramagnetic resonance (EPR) of $YBa_2Cu_3O_6$ compounds with $x \cong 6.0$ doped with $Dy^{3+}$, $Tb^{3+}$, and $Nd^{3+}$. The EPR spectra of $Dy^{3+}$ and $Tb^{3+}$ have been identified. The EPR of $Tb^{3+}$ is used also to study the effect of suppression of high $T_C$ superconductivity. The EPR of $Nd^{3+}$ is probably masked by the intense resonance of $Cu^{2+}$. All experimental EPR results compare well with theoretical estimations.

## 1. Introduction

$YBa_2Cu_3O_x$ (YBaCuO) with x > 6.35 is a superconductor ($T_c \approx 92$ K at $x \cong 7.0$) which has been studied by different techniques including methods of magnetic resonance. Among them until recently nuclear magnetic resonance (NMR) was by far dominant. The application of electron paramagnetic resonance (EPR) to high $T_C$ cuprates was restricted by the problem of EPR silence in these compounds [1]. The EPR line broadening estimated from the spin–phonon interaction ($\cong 10^{10}$ Hz) [2] is too large to detect the spectrum with EPR spectrometers working at the usual frequencies. Strong EPR signals observed at g ≈ (2–2.2) on the nominally pure high $T_C$ cuprates either at T < 40 K [3, 4, 5] or at higher temperatures [6, 7] were probably due to impurity phases or result from atmospheric degradation [1]. A further approach for EPR studies of high $T_C$ cuprates is to dope these compounds with small amounts of some paramagnetic ions to probe magnetic spin susceptibility and crystal fields of the $CuO_2$ bilayers. The EPR of rare-earth ($R^{3+}$) such as $Gd^{3+}$ [8, 9, 10], $Er^{3+}$ and $Yb^{3+}$ [11, 12, 13] has been used to investigate the intrinsic behavior of YBaCuO. With this technique it is possible to obtain data on the superconducting gap and pseudo-gap similar to those measured by inelastic neutron scattering and NMR. The corresponding *g*-values of rare-earth ions derived from EPR experiments can be used to determine exactly the parameters of the crystalline electric field (CEF) which are usually extracted from inelastic neutron scattering studies. The knowledge of *g*-values and corresponding magnetic moments allows to estimate dipole–dipole and exchange interactions both in diluted and concentrated compounds. The advantage of the EPR technique in this context is its time domain of observation being two to three orders of magnitude shorter than that of NMR. In the present work we extend the EPR investigations to some other $R^{3+}$-ions ($Dy^{3+}$, $Tb^{3+}$, $Nd^{3+}$) in order to clarify their widely discussed real valent states and positions in YBaCuO.

## 2. Sample preparation

We have studied polycrystalline samples of $YBa_2Cu_3O_x$ with a low oxygen content ($x \cong 6.0$), prepared using a conventional solid state reaction. Starting materials (powders of $Y_2O_3$, $BaCO_3$, and $CuO$) were dried at T ~ 400-500°C, mixed in appropriate stoichiometric amounts, and then milled into powder. $R^{3+}$- dopants were added using oxides $R_2O_3$ in a ratio R:Y=1:100. These mixtures were converted to $RYBa_2Cu_3O_7$ by thermal treatment and then the oxygen content was reduced to the specified value $x$, by heating the samples at a rate of 10 °C/min to 600–850 °C in a vacuum furnace. The oxygen deficiency obtained in this way was calculated from sample mass and the known data of the furnace (volume, temperature, and increase of pressure) using the ideal gas law. Subsequently it was adjusted by the reabsorption of the oxygen released into the furnace upon slow cooling (1 °C/min). The exact value of oxygen content $x$ depends on the annealing procedure and was determined with an accuracy of ±0.04.

## 3. Experimental details

The EPR behavior of YBaCuO compounds is strongly anisotropic, therefore both non-oriented polycrystalline samples and so-called "quasi-crystals" were investigated in this work. The powder samples were ground and mixed with either paraffin or epoxy resin. To prepare the "quasi-crystals," these samples were placed in a strong magnetic field ($\geqslant$ 15 kG) until the resulting suspension hardened. As a result of this treatment the $c$ axis of the individual crystallites should be aligned along the direction of the external magnetic field. EPR measurements were carried out on a IRES-1003 X-band ($\approx$ 9.5 GHz) spectrometer in the temperature range from 4 to 100 K and magnetic field up to 7 kG.

## 4. Results and discussion

### 4.1. Dysprosium

The EPR spectrum of a non-oriented sample $Dy_{0.01}Y_{0.99}Ba_2Cu_3O_x$ shows a very intense line with g $\approx$ 2 and three other broad signals in the magnetic field range from 400 to 1500 G with $g$-factors $g_A \approx 11.5\pm1.5$; $g_{B1} \approx 7.0\pm1.0$, and $g_{B2} \approx 4.5\pm1.0$ (Fig. 1). We assign the three resonances at low field to $Dy^{3+}$.

The temperature dependence of the intensity of line *A* obeys the Curie-law. The influence of parameters other than temperature has been eliminated by normalizing to the signal at g ≈ 2. The intensity of resonances marked *B* increases with temperature, pointing to the fact that line *A* stems from the ground state and lines *B* originate from the first excited state of the $Dy^{3+}$ ion. According to inelastic scattering experiments [14] a first excited CEF energy level of $Dy^{3+}$ in $DyBa_2Cu_3O_7$ is situated at ≈27 cm$^{-1}$≡38 K and therefore gets also populated at T > 10 K. The EPR spectrum of the quasi-single crystal sample for H∥C is very similar to that of the non-oriented sample. This means that only a small part of individual "crystallites" prepared were aligned in the external magnetic field. Nevertheless, the EPR spectrum in the perpendicular orientation of the magnetic field (H⊥C) differs from the parallel case—the *A* line almost disappears and within the *B* group the intensity of the line at $g_{B2}$ decreases with respect to that at $g_{B1}$. Although these changes are rather small we can conclude that the *A* line corresponds to $g_\parallel$ of the ground state doublet, and $g_{B2}≡g_\parallel$, $g_{B1}≡g_\perp$ are the principal *g* values of the excited doublet.

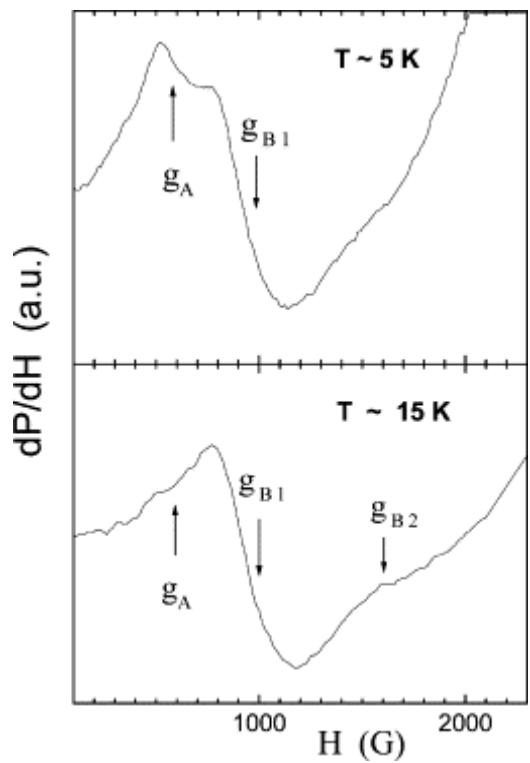

Fig. 1. EPR spectra of the non-oriented sample $Dy_{0.01}Y_{0.99}Ba_2Cu_3O_6$ at T = 5 K (upper panel) and T = 15 K (lower panel).

The EPR spectrum of the excited Kramers doublet is rarely observed. For example, the EPR of $Er^{3+}$ ions in $CaWo_4$ crystals has been attributed to the contribution of excited CEF energy levels [15], and the energy of the first excited doublet of $Er^{3+}$ (8.5–18.0 cm$^{-1}$) is very close to that of $Dy^{3+}$ in YBaCuO.

The very intense EPR line at g ≈ 2 could not prevent the observation of the EPR signal on $Dy^{3+}$ ions because of a sufficiently large value of *g*-factor.

Regarding the large linewidth, one first has to take into consideration that $Dy^{3+}$ like most rare-earth ions, undergoes a strong spin–phonon interaction leading to a decrease of relaxation times with a high power of *T*. In addition, due to the close lying first excited level the Orbach–Aminov relaxation process with its exponential temperature dependence should be effective. This suggestion can be confirmed by the Mössbauer spectroscopy data of the $Dy^{3+}$ paramagnetic relaxation rates in $DyBa_2Cu_3O_{7-x}$ [16], which reveal the Orbach–Aminov process as the dominant phonon driven relaxation mechanism above 5 K. The corresponding contributions to the EPR linewidth (due to the spin–lattice relaxation rate of $3 \cdot 10^9 s^{-1}$ at T = 5 K) can be estimated as ≈30 G (for g = 11.5) and ≈50 G (for g = 7.0). However, the drastic broadening of the EPR linewidth at T > 5 K and its disappearance at T ≈10 K ($1/T_1 \approx 10^{10} s^{-1}$) has not been observed using the EPR method. Probably, the small concentration of $Dy^{3+}$ ions (≈1%) caused this behavior in our samples in contrast to a very effective spin–lattice relaxation in the concentrated $DyBa_2Cu_3O_{7-x}$ compound.

The absence of any change with temperature shows that up to T = 20 K the experimentally observed linewidth is not influenced by the spin–lattice relaxation, but inhomogeneously broadened by variations of the crystalline electric field acting on $Dy^{3+}$ ions at different sites. One can neglect the other broadening interactions (such as dipole–dipole, exchange, etc.) because of the small concentration of $Dy^{3+}$.

Stark splitting and *g*-values were calculated by using crystal field Hamiltonian

$$H_{CEF} = \sum_{n=1}^{3} \sum_{m=0,2} \gamma_{2n} A_{2n}^{2m} O_{2n}^{2m} \qquad (1)$$

which describes the tetragonal crystal field ($\gamma_{2n}$ are the Stevens coefficients, $O_{2n}^{2m}$ are the Stevens Operators and $A_{2n}^{2m}$ are the corresponding CEF parameters.) Using this Hamiltonian with the CEF parameters appropriate for $Er^{3+}$ in ErBaCuO with the oxygen index *x* = 6.09 [17], we calculated for the first three crystal field levels of the $^6H_{15/2}$ ground term of $Dy^{3+}$ relative energies of 0, 11.6, and 39.5 $cm^{-1}$ and $g_\| = 11.5$ and $g_\perp = 0.54$ for the lowest, $g_\| = 4.6$ and $g_\perp = 8.85$ for the first excited Kramers doublet.

We can use the data for $DyBa_2Cu_3O_7$ [14] as CEF parameters supposing that the rhombic parameters $A_2^2 = A_4^2 = A_6^2 = A_6^6 = 0$. The standard calculations yield that the three lowest CEF energy levels are situated at 0, 30.2, 44.2 cm$^{-1}$, respectively, and the *g*-factor values of the ground state are $g_{\parallel}$ = 14.37 and $g_{\perp}$ = 0.11, and those of the first excited state are $g_{\parallel}$ = 7.9, $g_{\perp}$ = 7.03. We can also obtain an another set of similar data assuming that $A_2^0$ becomes two times smaller by alteration of the oxygen content from 7 to 6, just as the observed case for Er$^{3+}$ in ErBaCuO [17]. There the three lowest CEF energy levels are lying at 0, 14.1, 58.4 cm$^{-1}$, respectively, $g_{\parallel}$ = 14.15, $g_{\perp}$ = 0.027 (ground state), and $g_{\parallel}$ = 6.8, $g_{\perp}$ = 7.8 (first excited state). A comparison of these estimations with experimental results and corresponding literature data shows that the calculations depend strongly on the value of CEF parameters, especially on the value of $A_2^0$. The *g*-factor values derived using the CEF parameters for Er$^{3+}$ in ErBaCuO [17] are very close to the experimental data for the excited doublet. We do not possess any experimental data about CEF level structure on Dy$^{3+}$ ions in samples with $x \approx 6.0$ K. Although this structure changes only slightly during the degradation of oxygen content from 7 to 6 on Er$^{3+}$ and Nd$^{3+}$ ions, such alterations can be more than , 10 cm$^{-1}$ [17, 18].

Notice that $g_{\perp}$ < 1 for the ground state. However, we could not define the exact value because our EPR measurements have been performed in magnetic fields below 7 kG. We have also investigated the EPR behavior of Dy$^{3+}$ in YBa$_2$Cu$_3$O$_x$ with $x$ = 6.1, 6.46, and 6.96. In the first case, the EPR spectrum is similar to that for $x$ = 6.0. For the other two samples the typical low field absorption of superconducting compounds prevented the observation of EPR. Likodimos et al. [19] reported a broad asymmetric EPR signal in several polycrystalline DyBa$_2$Cu$_3$O$_x$ samples. They ascribed this resonance to the ground state of Dy$^{3+}$ ions which due to the high concentration experience a strong line broadening by dipolar interaction. In our opinion, role of the first excited state should be also taken into account in the theoretical estimations.

**4.2. Terbium**

RBa$_2$Cu$_3$O$_x$ compounds (R–Y, rare-earth) are superconducting below 92 K, except for R=Ce, Pr, Tb. The reasons for such suppression were discussed previously [20], and with respect to Tb it is often pointed out that this ion has a stable 4+ state. The presence of Tb$^{4+}$ might explain why TbBa$_2$Cu$_3$O$_7$ does not form. The direct determination of the oxidation state and hybridization of Tb in the RBa$_2$Cu$_3$O$_x$ ( 6 < x < 7) series is therefore expected to provide information for the understanding of the ability of selected rare-earth ions to suppress superconductivity. Both valence states of terbium, Tb$^{3+}$ and Tb$^{4+}$ can be detected by EPR. However, their EPR spectra are rather different [21, 22, 23]. The electronic configuration of Tb$^{4+}$ (an ion with a half-filled 4*f*-shell like Gd$^{3+}$) is 4$f^7$, and the ground state of the free ion is $^8S_{7/2}$.. The EPR spectrum of Tb$^{4+}$ consists of seven fine structure lines with a very small hyperfine constant *A* (of only few gauss) and can be observed even at room temperature. In contrast, EPR of the non-Kramers ion $Tb^{3+}(4f^8, {}^7F_6)$ has been observed only below 30 K. It is characterized by narrow resonance lines, $g = g_\parallel \approx 18$ and $A \approx 250$ G. The EPR spectrum of the sample $Tb_{0.01}Y_{0.99}Ba_2Cu_3O_6$ consists of a very intense EPR signal at $g \approx 2$ and several partly resolved lines at low magnetic field ($H < 700$ G). This low-field part is shown in Fig. 2a. It can be well interpreted in terms of the spin Hamiltonian for $Tb^{3+}$ ions [22]:

$$H = g_\parallel \beta H S_z + A S_z I_z + \Delta S_x, \quad (2)$$

with $S = \frac{1}{2}$, $I = 3/2$, $g_\parallel \approx 18$, and $A \approx 250$ G. Further evidence for the assignment of these low field lines to $Tb^{3+}$ is given by the fact that they could not been observed at $T > 30$ K. It is well known [21] that EPR of $Tb^{3+}$ ions are not observed at $T > 30$ K because of fast spin–lattice relaxation. In principle, EPR measurements at different frequencies would be necessary for a precise determination of $g_\parallel$. Calculations using the CEF parameters for $Er^{3+}$ in $ErBa_2Cu_3O_{6.09}$ [17] however give already a very reasonable value of 17.8 for this parameter. The zero field splitting $\Delta$ between the lowest singlet levels involved in the resonance can be estimated from the experimental results by converting Eq. (2) to the form [22]

$$h\nu = [(g_\parallel \beta H + Am)^2 + \Delta^2]^{1/2} \quad (3)$$

for the allowed transitions with $\Delta M = \pm 1$, $\Delta m = 0$. This limits the lowest frequency ($\nu$) at which EPR can be observed to $\nu = \Delta/h$, and sets by our microwave frequency $\nu \approx 9.5$ GHz an upper limit of 0.32 $cm^{-1}$ for $\Delta$. This value can be improved by an analysis of the hyperfine structure in Fig. 2a, yielding $A = 260$ G $\equiv 6.47$ GHz and the coincidence of the resonances with $m = -3/2$ and $m = -1/2$ at a field of approximately 150 G (cf. Fig. 2b).

From this coincidence follows $h\nu = (A^2 + \Delta^2)^{1/2}$ [22] and (within the accuracy of the measurement) $\Delta = 0.23$ $cm^{-1} \equiv 6.9$ GHz. The determination of such a small splitting shows the power of EPR, since it could not be obtained by other methods such inelastic neutron scattering. Theoretical estimates [20] on the other hand report $\Delta = 0.005$ meV $\equiv 0.04$ $cm^{-1}$, i.e., a value one order of magnitude smaller than our experimental data. A statement about the presence $Tb^{4+}$ ions gets complicated by the very intense signal at $g \approx 2$ (cf. Section 4.3). Within this limitation no evidence for the existence of Tb in the 4+ state could be found.

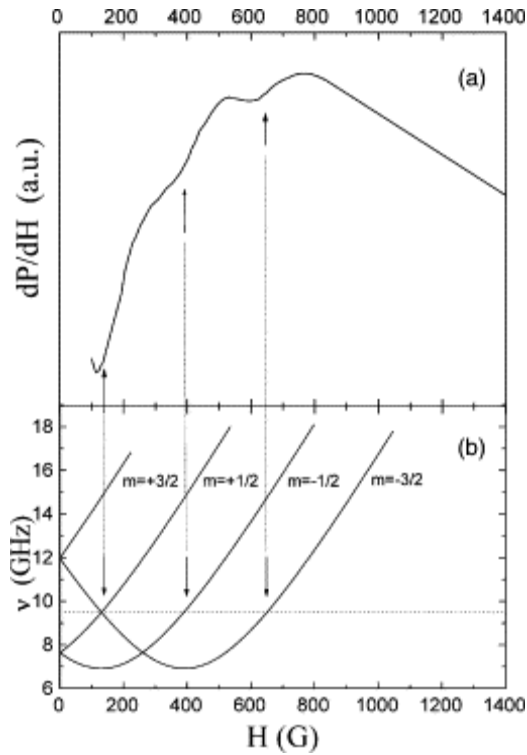

Fig. 2
*a*. EPR signal of $Tb^{3+}$ ions in $Tb_{0.01}Y_{0.99}Ba_2Cu_3O_6$ measured at T = 7.6 K and at the frequency of 9.5 GHz.
*b*. The variation of the resonance frequencies of the hyperfine components of $Tb^{3+}$ ions in $Tb_{0.01}Y_{0.99}Ba_2Cu_3O_6$ in the low-magnetic field region. Dashed line corresponds to the frequency of 9.5 GHz; arrows indicate resonance fields at the frequency ~9.5 GHz.

**4.3. Neodymium**

No EPR studies were reported for YBaCuO compounds with a low concentration of $Nd^{3+}$. There are only a few publications devoted to EPR in (Y,Nd)BaCuO, where either all or 50% of $Y^{3+}$ ions were replaced by $Nd^{3+}$ [4, 24, 25]. Two EPR signals were observed in the oxygen deficient non-superconducting NdBaCuO at 77 K [24], a narrow one ($\Delta H_{pp} \approx 150$ G) at g = 3.6 and a broad one ($\Delta H_{pp} \approx 1000$ G) at g = 2.13. Both were ascribed to $Nd^{3+}$. However, it is well known that EPR cannot be detected at liquid nitrogen temperature because of a very fast spin–lattice relaxation of neodymium [18]. The authors of [25] investigated the EPR of $Nd_{0.5}Y_{0.5}Ba_2Cu_3O_x$ compounds in the temperature range 3.6–70 K and concluded that the absence of the $Nd^{3+}$ resonance is due to very fast spin–lattice relaxation. Polycrystalline and single crystal samples of NdBaCuO were also investigated at frequencies 9.3 and 35 GHz in the temperature range from 4 to 300 K by Baranov et al. [4]. The observed intense and broad EPR signal ($g_\parallel \approx 2.15$; $g_\perp \approx 2.2$ at T = 10 K) was ascribed to $Cu^{2+}$ ions and not to $Nd^{3+}$. Thus, no EPR of $Nd^{3+}$ in YBaCuO has been observed until now. Our EPR experiments on the $Nd_{0.01}Y_{0.99}Ba_2Cu_3O_6$ revealed only a very intense EPR line at g ≈ 2 and no other signals, confirming the above. The observed spectrum most likely indicates only the presence of $Cu^{2+}$ ions ($g_\parallel \approx 2.2$ and $g_\perp \approx 2.0$). We estimate the *g*-factors of $Nd^{3+}$ theoretically in order to find the reason for the absence of the $Nd^{3+}$ EPR. The CEF parameters of $Nd^{3+}$ in $NdBa_2Cu_3O_x$ used in our calculations describe quite well the CEF level structure of all multiplets multiplets $^4I_{9/2}$, $^4I_{11/2}$, $^4I_{13/2}$, and $^4I_{15/2}$ [26]. They yield $g_\parallel \approx 2.33$ and $g_\perp \approx 2.56$. These values are very close to the *g*-factor of $Cu^{2+}$ ions in YBaCuO. Therefore, the EPR signal of $Nd^{3+}$ in YBaCuO can easily be masked by the very intense resonance of $Cu^{2+}$ occurring simultaneously in the same magnetic field range.

There is no any evidence of EPR of the first six lanthanides in YBCO in the literature. Thus, the reason for the absence of the $Nd^{3+}$ EPR may lie deeper. In this connection it should be interesting to investigate this compound with a low concentration of cerium, praseodymium, and samarium ions. The obvious difficulties are the small values of *g*-tensor of $Sm^{3+}$ (EPR might be observed at extremely high fields) and singlet ground state of $Pr^{3+}$ non-Kramers ions. Thereby we expect that EPR of $Pr^{4+}$ ions might be only detected.

**5. Conclusions**

We have performed EPR studies on $YBa_2Cu_3O_6$ samples doped with 1% of three different rare-earth ions ($Dy^{3+}$, $Tb^{3+}$, and $Nd^{3+}$). The EPR of $Dy^{3+}$ has been identified both within the lowest ($g_\parallel \approx 11.5$) and the first excited ($g_\parallel = 4.2$ and $g_\perp \approx 7.0$) doublet. The EPR of $Tb^{3+}$ was detected within the lowest singlet states separated by $\Delta = 0.23$ cm$^{-1}$ ($g_\parallel \approx 17.8$; $A = 260$ G $\equiv 6.47$ GHz). No $Tb^{4+}$ EPR resonance was found. Within the experimental limitations this seems to confirm that the reducing of $T_C$ in Tb-doped $YBa_2Cu_3O_x$ superconductors and the impossibility to prepare $TbBa_2Cu_3O_7$ are not related to the existence of $Tb^{4+}$ ions. The experimental EPR results on $Dy^{3+}$ and $Tb^{3+}$ can be well described by corresponding spin Hamiltonians of appropriate symmetry. We note in addition that the EPR, signal of $Nd^{3+}$ is most likely masked by the very intense EPR of $Cu^{2+}$ ions in YBaCuO.

**Acknowledgements**

The authors are grateful to Dr. A. Hofstaetter (Giessen, Germany) for help with the revision of the paper. This work was supported by the Swiss National Science Foundation (Grant No. 7SUPJ062258), by the CRDF (grant REC-007) and by the Russian Program "Universities of Russia" (Grant No. 991327).

marat.gafurov@ksu.ru